\documentclass[aps,prl,twocolumn,preprintnumbers,
               showpacs,nofootinbib]{revtex4}
\newcommand{\PRE}[1]{}       

\usepackage{bm}
\usepackage{epsfig}
\newcommand{\postscript}[2]{\setlength{\epsfxsize}{#2\hsize}
   \centerline{\epsfbox{#1}}}

\newcommand{\eqref}[1]{Eq.~(\ref{#1})}

\begin{document}

\preprint{CLNS 07/1996}

\title{
\PRE{\vspace*{1.5in}}
Extremely Long-Lived Charged Massive Particles as A Probe for 
Reheating of the Universe
\PRE{\vspace*{0.3in}}
}

\author{Fumihiro Takayama%
\PRE{\vspace*{.2in}}
}
\affiliation{
Institute for High Energy Phenomenology, 
Cornell University, Ithaca, New York 14853, USA
\PRE{\vspace*{.1in}}
}


\begin{abstract}
\PRE{\vspace*{.1in}}
We discuss the impact of charged massive particle big bang nucleosynthesis(CBBN) 
to explore the nature of the reheating of the Universe in the case that 
a new extremely long-lived charged massive particle(CHAMP) exists. 
If the mass of the CHAMP is within collider reach and it's lifetime is 
longer than $10^4$sec, the comparison between the CBBN prediction and observed $^6$Li abundances 
may indicate nonstandard reheating in the early Universe without relying on details of the decay 
properties. Even if the CHAMP mass is outside the reach of colliders, the cosmological considerations 
may provide a nontrivial hint for the existence of such very heavy long-lived CHAMPs from 
the late Universe if the daughter particles are the dominant component of the 
present dark matter. We consider a low reheating temperature model as an example 
of the nonstandard reheating scenarios. 
\end{abstract}


\maketitle

\section{Introduction}
We are living in an interesting time for particle physics. The 
CERN Large Hadron Collider (LHC) experiment will start this year and 
may show exciting results soon. 
 Also recent significant developments of cosmological and astrophysical 
observations have improved our knowledge of particle physics.

The fact that the large fraction of matter in the Universe 
is not accounted for by standard model particles motivates further 
considerations for the role of hypothetical particles in the early 
Universe and the connection to collider physics. Recent theoretical 
studies have shown the possibility of extremely long-lived charged 
massive particles(CHAMPs)~\cite{Drees:1990yw,slepton,Ellis:2003dn,
Feng:2004mt}. 
The search for CHAMPs and the determination of it's properties may be 
an exciting challenge for future collider experiments and cosmological 
observations. 

Under continuous efforts in the past decade, the standard big bang 
nucleosynthesis(SBBN) theory has 
been well established to precisely predict the primordial light element 
abundances~\cite{Coc:2003ce} and may be used as a probe to find or constrain 
new physics beyond the standard model in the late time Universe
~\cite{hadronicBBN:2004qu,Jedamzik:2004er,
Kohri:2006cn,Pospelov:2006sc,Kaplinghat:2006qr,Cyburt:2006uv,
Hamaguchi:2007mp,Bird:2007ge,Kawasaki:2007xb}. 
As recent papers have pointed out~\cite{Kohri:2006cn,Pospelov:2006sc,
Kaplinghat:2006qr,Cyburt:2006uv,Hamaguchi:2007mp,Bird:2007ge,Kawasaki:2007xb}, 
if extremely long-lived CHAMPs existed in the early Universe, 
such a CHAMP might constitute a bound state with a light 
element after/during the BBN era and modify the SBBN prediction of light 
element abundances.\footnote{The bound state formation and the fate of stable 
CHAMPs had been also discussed in earlier works~\cite{DeRujula:1989fe}.} 
This might constrain the number density of long-lived 
CHAMPs depending on the lifetime.  

The cosmological constraints on thermally frozen long-lived CHAMPs 
have been discussed~\cite{Feng:2004mt,Ellis:2003dn,Steffen:2006hw}. 
In most of these papers, the standard radiation dominated Universe during 
the freeze-out of CHAMPs was assumed. 
In such a standard scenario, the relic density $\Omega_C \sim m_C^2$ and 
the number density $n_C\sim m_C$ where $m_C$ is
 the mass of the CHAMP, which implies severe constraint for heavier mass. 
 Recent analysis showed that the entire mass region may be disfavored by BBN 
if the lifetime is longer than $10^4$sec
~\cite{Pospelov:2006sc,Pradler:2006hh,Kawasaki:2007xb}. 
\footnote{For the lifetime $> 7\times 10^6$sec, this conclusion may be changed 
once CHAMPs are captured by protons~\cite{Kohri:2006cn}} 

On the other hand, above BBN constraints may not exclude the possibility of
the discovery of CHAMPs with lifetime $>10^4$sec. The violation of the above 
constraints does not necessarily mean contradictions but may indicate an 
evidence of additional new physics or lack of understanding of astrophysics. 
From this point of view, the importance of searching for extremely long-lived 
massive particles was also discussed in the context of the 
supersymmetric theory with a gravitino LSP ~\cite{Feng:2004mt}. 

In this paper, we discuss a possible impact of the discovery of 
extremely long-lived CHAMPs to explore the nature of the reheating of the Universe, 
where the standard scenario has assumed the radiation dominated Universe 
during and after the thermal freeze-out of the CHAMP relic density until the time 
when the present matter dominated the enegy of the Universe against the radiation. 
We focus on a low reheating temperature model as the possibility of 
the nonstandard Universe and extract constraints on the theoretical 
parameters. We calculate the relic density of CHAMPs frozen during/after 
reheating and estimate the CBBN prediction of primordial $^6$Li abundance 
by using a recent estimation of the nuclear reaction rate
~\cite{Hamaguchi:2007mp} and without assuming the Saha equation for bound 
state formation of a CHAMP and a $^4$He. 
We will find that the CBBN prediction of $^6$Li is sensitive to 
the number density of CHAMPs and the lifetime around $10^3-10^4$sec. 
If the lifetime is longer than $10^4$ sec, the comparison between observed $^6$Li 
and the CBBN prediction can impose a relevant constraint on the reheating temperature, 
which may potentially become an interesting restriction on some class of 
models of reheating after inflation or models of late time entropy production
~\cite{Endo:2006xg,Endo:2007cu,Kofman:2005yz}.
\footnote{Similar discussions to this paper may be interesting for 
colored or double charged particles~\cite{others}.}

\section{Low reheating temperature and the relic abundance of CHAMPs}
Here we consider the CHAMP production during/after reheating of the 
Universe. It is believed that the reheating of the Universe results 
from the thermalization of the decay products from the decay of 
coherent oscillations of a scalar field or the decay of some heavy massive 
particles such as moduli or gravitinos which do not thermalize with 
the thermal bath of SM particles after the energy dominated the Universe.   
 Here we consider reheating due to the decay of a scalar field $\phi$. 
For the case of a heavy massive particle such as moduli or gravitino,
 a similar discussion to this paper can be applied~\cite{Kohri:2005ru}. 
The reheating temperature is defined by
\begin{eqnarray}
\Gamma_{\phi}=\sqrt{\frac{4\pi^3 g_{\ast}(T_{\text{RH}})}{45}}
\frac{T_{\text{RH}}^2}{M_{\text{pl}}},
\end{eqnarray}
where $\Gamma_{\phi}$ is the decay rate of the scalar field $\phi$,
$g_{\ast}(T)$ is the number of massless degree of freedom at temperature $T$
 and $M_{\text{pl}}=1.2\times 10^{19}$GeV.
In this paper, we call this a $low$ reheating temperature scenario in which 
most of the decays of the scalar field $\phi$ occur after freeze-out of CHAMPs.
 The importance of particle productions during the reheating era and the 
difference from the standard radiation dominated Universe in thermally
 frozen nonrelativistic particles was discussed in
~\cite{Giudice:2000ex,Kohri:2005ru,Drees:2007kk}.
 
We estimate the relic density of heavy charged species as a function of 
the mass and reheating temperature.  
\footnote{The constraint on stable CHAMPs from heavy isotope searches was
 discussed in a similar set up~\cite{Kudo:2001ie}. On the other hand, notice that 
strictly speaking, the constraint from heavy isotope searches is constraining 
only the primordial abundance of stable CHAMPs which in general depends on a cosmological 
model of the early Universe. A heavy isotope constraint on stable CHAMPs produced 
by cosmic rays does not depend on cosmological models, but it still has model dependencies 
on the cosmic ray production cross section of CHAMPs ~\cite{Byrne:2002ri,Albuquerque:2003mi}.} 
We consider a simple reheating model described by the following equations
~\cite{Giudice:2000ex}.
\begin{eqnarray}
&&\frac{d \rho_{\phi}}{dt}+3H\rho_{\phi}=-\Gamma_{\phi}\rho_{\phi}
\nonumber\\
&&\frac{d \rho_R}{dt}+4H\rho_R=\Gamma_{\phi}\rho_{\phi}
+<\sigma v>2<E_C>[n_C^2-n_{\text{EQ}}^2]\nonumber\\
&&\frac{d n_C}{dt}+3Hn_C
=-<\sigma v>[n_C^2-n_{\text{EQ}}^2]
\end{eqnarray}
where $\rho_C=m_Cn_C$, $\rho_{\phi}$, $\rho_R$ 
are energy densities of the $\phi$ field and radiation, $H$ is the Hubble parameter, 
$H^2=(8\pi/3M_{\text{pl}}^2)(\rho_C+\rho_{\phi}+\rho_R)$ and 
$<E_C>=\sqrt{m_C^2+(3T)^2}$. The kinetic equilibrium for CHAMPs and radiations is assumed. 
Also we assumed that CHAMPs annihilate into radiation or the particles in the final state rapidly 
convert into radiation.
As the initial condition, we use $\rho_{\phi}=(3/8\pi)M_{\text{pl}}^2
H_I^2$, $\rho_R=\rho_C=0$. The initial Hubble parameter $H_I$ is taken 
significantly early before the completion of reheating at $\Gamma_{\phi}^{-1}$ 
and the freeze-out of CHAMPs. 

In this paper, we do not consider more complicated cases
 of reheating after inflation and assume that this is the last reheating of 
the Universe. But notice that $\rho_R=\rho_C=0$ may not always be satisfied 
if this reheating occurs after the Universe went through radiation 
dominated once as a result of earlier reheating~\cite{Pradler:2006hh}.   

Here notice that we have assumed two things. First, we assumed that the branching ratio 
of the $\phi$ decay into CHAMPs and the secondary production of CHAMPs due to
 the $\phi$ decay products are negligible during reheating. If the branching ratio is not small enough, 
we have to take into account the additional nonthermal contribution to 
the CHAMP relic after CHAMPs freeze-out~\cite{Kohri:2005ru}. The simple solution
 would be $m_{\phi}<m_C$ and the other possibility was also discussed in \cite{Endo:2007cu}. 
Second, we do not consider the significant amount of decays of $\phi$ 
directly into hidden particles like gravitinos which has a very 
small coupling with standard model particles, that is, we assume that such a
 component of hidden particle does not exceed the relic density of present dark matter and 
does not overclose the Universe. 

We are interested in the production of nonrelativistic particles 
during/after reheating. 
In this paper, we will use the analytical solution for thermally frozen 
CHAMP relic density of the above differential equations~\cite{Giudice:2000ex}. 

 The particle freeze out during reheating may be classified into two cases, 
the case that chemical equilibrium is not established before 
freeze-out, and the case that chemical equilibrium is established. 
The annihilation cross section determines which of the two cases applies. 
Since we are interested in the production of CHAMPs, the minimal coupling would 
be the coupling with a photon. Assuming {\it s}-wave processes dominates,
 we have
\begin{eqnarray}
<\sigma_{\text{ann}}v>=\frac{4\pi\alpha^2}{m_X^2}\gamma
\end{eqnarray}
where we take $\gamma=1$.
Chemical equilibrium during reheating era is established before freeze-out if
\begin{eqnarray}
\frac{T_{\text{RH}}}{m_C}&>&
1\times 10^{-6}\nonumber\\
&&\times \gamma^{-1/2}
\sqrt{\frac{2}{g}[\frac{g_{\ast}(T_{\ast})}{g_{\ast}(T_{\text{RH}})^{1/2}}]}
[\frac{m_C}{10^2\text{GeV}}]^{1/2}
\end{eqnarray}
This condition follows from $n_C^{\text{Max}}>n_{\text{eq}}(T_{\ast})$ 
where $T_{\ast}=4m_C/17$ for {\it s}-wave corresponds to the temperature at which 
most of the production takes place~\cite{Giudice:2000ex}. 
Notice that for the thermal freeze-out during reheating, the typical 
freeze-out temperature $T_F$ may be $O(0.1)\times m_C$. If $T_F>T_{\text{RH}}$ 
and the above condition is satisfied, the chemical equilibrium in CHAMPs will 
be established and the relic will freeze out during reheating, for example, 
for $T_{\text{RH}}/m_C\sim O(10^{-3})$ and $m_C < 10^7$GeV. 

For the nonchemical equilibrium production, the present time 
relic density of CHAMPs is
\begin{eqnarray}
\Omega_C^{NT(Low)}h^2=
0.13\times
(\frac{g}{2})^2[\frac{g_{\ast}(T_{\text{RH}})^{1/2}}{g_{\ast}(T_{\ast})}]
^3[\frac{10^3 T_{\text{RH}}}{m_C}]^7\gamma
\end{eqnarray}

For the chemical equilibrium production, the final relic 
abundance is approximately given by $<\sigma_{\text{ann}} v> n_C^{eq}
=H(T_F)$ and by taking account of the dilution of the frozen relic until
 the reheating completes. The Hubble parameter during the reheating is 
$H=\sqrt{5\pi^3g_{\ast}(T)^2/9g_{\ast}(T_{\text{RH}})}T^4/T_{\text{RH}}^2
M_{\text{pl}}$~\cite{Giudice:2000ex}. 
Notice that $T\sim a^{-3/8}$ during reheating,
 so that $a(T_{\text{RH}})^3/a(T_F)^3=(T_F/T_{\text{RH}})^8$
~\cite{Giudice:2000ex}. Then,
\begin{eqnarray}
\Omega_C^{TH(Low)} h^2=3.3\times 10^{-8}
[\frac{g_{\ast}(T_{\text{RH}})^{1/2}}{g_{\ast}(T_F)}]
\frac{T_{\text{RH}}^3\text{GeV}^{-2}}{\gamma m_C x_F^{-4}}
\end{eqnarray}
where $x_F=m_C/T_F$.
 The freeze-out temperature is fixed by the solution of $<\sigma_{\text{ann}}v>
n_C^{eq}(T_F)=H(T_F)$.
\begin{eqnarray}
x_F=\ln[\frac{6\pi g}{\sqrt{10}\pi^2}
\frac{g_{\ast}(T_{\text{RH}})^{1/2}}{g_{\ast}(T_F)}
\frac{M_{\text{pl}}T_{\text{RH}}^2}{m_C^3}\alpha^2\gamma x_F^{5/2}]
\end{eqnarray}
 Here we implicitly assume that the maximum temperature is higher than the
freeze out temperature, which may constrain the initial energy density 
of the $\phi$ field.
This case reduces to the known thermal freeze-out formula
 within a few  10 percent differences if we take the limit 
$T_{\text{RH}}= T_F$. That is, in this limit, eliminating dilution,
\begin{eqnarray}
\Omega_C^{TH} h^2\simeq 4.2\times 10^{-8}
[\frac{1}{g_{\ast}(T_F)^{1/2}}]
\frac{m_C^2\text{GeV}^{-2}}{\gamma x_F^{-1}}
\end{eqnarray}
\begin{eqnarray}
x_F=\ln[\frac{3\sqrt{5} g}{\sqrt{2}\pi^2}
\frac{1}{g_{\ast}(T_F)^{1/2}}
\frac{M_{\text{pl}}}{m_C}\alpha^2\gamma x_F^{1/2}]
\end{eqnarray}
This gives roughly $n_C/n_H\sim 3\times 10^{-4}(m_C/100\text{GeV})/\gamma$.
 
\section{CHAMP Big Bang Nucleosynthesis (CBBN)}
Next we consider the impact of CBBN for the number density of CHAMPs.
Recently it has been pointed out that the bound state formation of 
a CHAMP and a light element at lower temperature of the binding energy 
may change the nuclear reaction rates of BBN and eventually change 
the light elements abundance from SBBN~\cite{Kohri:2006cn,Pospelov:2006sc,
Cyburt:2006uv,Bird:2007ge}. 
Especially Pospelov pointed out that because of the strong 
electro-magnetic field due to the bound CHAMP, CBBN may have significant 
changes in the nuclear reaction rates from SBBN's for radiative processes 
which are highly suppressed in SBBN~\cite{Pospelov:2006sc}. 
The CBBN nuclear reaction rate for $^6$Li production has recently been 
evaluated by the use of the state-of-the-art coupled channel method
~\cite{Hamaguchi:2007mp}. 
We use the reaction rate here. for $T<100$keV,
\begin{eqnarray}
&&<\sigma_{^6\text{Li}}^{CBBN}v>
=3.4\times 10\times\text{GeV}^{-2}\nonumber\\
&&~~~\times (1-0.34(\frac{T}{10^9\text{K}}))
(\frac{T}{10^9\text{K}})^{-2/3}e^{-5.33(\frac{T}{10^9\text{K}})^{1/3}}
\end{eqnarray}

In this paper, we do not include decay effects and only consider
 effects of the CBBN on $^6$Li. Actually in the case of sufficiently 
low abundances compared to the present dark matter relic density, the impact on 
$^6$Li due to the decays will not be significant~\cite{hadronicBBN:2004qu}. 
In this paper, we consider the case where the number density is smaller 
than $^4$He but larger than other light elements. 
Then we can simplify the discussion of the capture of CHAMPs, that is, 
essentially the important part is only the capture by $^4$He. 
Since elements heavier than $^4$He are extremely rare, if the number 
density $n_C/n_p>10^{-10}$, we can safely ignore the captures by 
elements other than $^4$He to evaluate the number density of the bound state 
($C$,$^4$He). 
\footnote{The formation of the bound state with heavier elements than $^4$He 
uses only a small fraction of CHAMP. But if the number density is not small 
relative to $^4$He abundance, the bound state formation with heavier elements 
might have some impact on evaluating the primordial abundance of such heavier 
elements through CBBN processes, for example, 
in $^7$Li, etc.~\cite{Kohri:2006cn,Bird:2007ge}}.
After the capture by $^4$He starts, the probability of CHAMP captures by
 elements lighter than $^4$He is extremely small and can be safely 
ignored until the capture of protons starts 
at $T=0.6$keV ($t=7.2\times 10^6$ sec). 
Once proton capture starts, the remaining free CHAMPs will be immediately captured 
by protons. In this paper, we do not consider the effect of the bound state 
with a proton which may have some impact on light element abundance 
including $^6$Li. The approximation will be safe for the case of 
$\tau_C< 7\times 10^6$sec.
\footnote{However, the bound state formation with protons
 may decrease $^6$Li through $^6\text{Li}((C,p),\alpha)^3\text{He}$
~\cite{Kohri:2006cn}. But the estimation assumed that the nuclear reaction 
does not have Coulomb suppression and the stability of the bound state
 during nuclear reaction, which may not be a good approximation 
for the case of proton captures because the Bohr radius of the bound state is 
large relative to typical nuclear size. For the other bound sate with D, T, 
etc. would only contribute the small production of $^6$Li.}

Assuming that the reheating temperature is high 
enough for thermal BBN to occur, we solve following equations.
\begin{eqnarray}
&&\frac{d n_{^4\text{He}}}{dt}+3Hn_{^4\text{He}}=\nonumber\\
&&~~~~~~~-<\sigma_{\text{rec}}v>
(n_Cn_{^4\text{He}}-n_{(C,^4\text{He})}\tilde{n}_{\gamma})
+\frac{1}{\tau_C}n_{(C,^4\text{He})}\nonumber\\
&&\frac{d n_C}{dt}+3Hn_C=
\nonumber\\
&&~~~~~~~-<\sigma_{\text{rec}}v>
(n_Cn_{^4\text{He}}-n_{(C,^4\text{He})}\tilde{n}_{\gamma})\nonumber\\
&&~~~~~~~~~~~~~~~+<\sigma_{^6\text{Li}}^{CBBN}v>n_{(C,^4\text{He})}n_D
-\frac{1}{\tau_C}n_C
\nonumber\\
&&\frac{d n_{(C,^4\text{He})}}{dt}+3Hn_{(C,^4\text{He})}=\nonumber\\
&&~~~~~~~<\sigma_{\text{rec}}v>
(n_Cn_{^4\text{He}}-n_{(C,^4\text{He})}\tilde{n}_{\gamma})
\nonumber\\
&&~~~~~~~~~~~~~~~-<\sigma_{^6\text{Li}}^{CBBN}v>n_{(C,^4\text{He})}n_D
-\frac{1}{\tau_C}n_{(C,^4\text{He})}\nonumber\\
&&\frac{d n_{^6\text{Li}}}{dt}+3Hn_{^6\text{Li}}=
<\sigma_{^6\text{Li}}^{CBBN}v>n_{(C,^4\text{He})}n_{\text{D}}\nonumber\\
&&\frac{d n_{\text{D}}}{dt}+3Hn_{\text{D}}=
-<\sigma_{^6\text{Li}}^{CBBN}v>n_{(C,^4\text{He})}n_{\text{D}}
\end{eqnarray}
where $n_{\gamma}=2\zeta(3)T^3/\pi^2$,
\begin{eqnarray}
\tilde{n}_{\gamma}=n_{\gamma}\frac{\pi^2}{2\zeta(3)}(\frac{m_{^4\text{He}}}
{2\pi T})^{3/2}e^{-E_{\text{bin}}/T}
\end{eqnarray}
where we take the binding energy of $^4$He to a CHAMP $E_{\text{bin}}=$311keV
~\cite{DeRujula:1989fe}.
We use the following recombination cross section:
\begin{eqnarray}
<\sigma_{\text{rec}}v>
=\frac{2^9\pi\alpha Z_{^4\text{He}}^2\sqrt{2\pi}}{2e^4}
\frac{\tilde{E}_{\text{bin}}}{m_{^4\text{He}}^2\sqrt{m_{^4\text{He}} T}}
\end{eqnarray} 
where $\tilde{E}_{\text{bin}}=\alpha^2Z_C^2Z_{^4\text{He}}^2
m_{^4\text{He}}/2$~\cite{DeRujula:1989fe}. 
\footnote{In this estimation of recombination cross section, 
a point particle for $^4$He is assumed and only capture into the 1S state is 
considered. The estimation may have uncertainties from this assumption, 
the ignorance of the internal structure of nuclei and the contributions 
from captures into higher level, which directly reflct to the CBBN prediction.}
For a nonthermal photon emitted in the recombination process, we assume the
 rapid thermalization which would be valid in the deeply radiation dominated 
Universe, that is, we always assume thermal distribution for photon and 
also kinetic equilibrium for all light elements. 

We evaluate $^6$Li abundance by use of Eq.(11). We assume 
 the Saha-type equation $n_{(C,\text{He})}=n_Cn_{^4\text{He}}/\tilde{n}_{\gamma}$
and $n_C=n_C^{in}e^{-(t/\tau_C)}$ for the number density of the bound state 
until $T=10$ keV, and connect to the set of differential 
equations of Eq.(11) for temperature below 10 keV and numerically solved them. 
 At $T>$10 keV, since the photo-destruction rate $<\sigma_{\text{rec}}v>
\tilde{n}_{\gamma}/H> 10^2$ and the bound state formation rate 
is also larger than the expansion rate of the Universe in the third equation 
of eq(11), the use of the Saha equation would be a good approximation.
 Failure of the Saha approximation may happen if
$3H\sim <\sigma_{\text{ann}}v>\tilde{n}_{\gamma}$, when the rate of 
bound state formation is faster than the expansion rate of the Universe.
We also used a SBBN value for the light elements as the 
initial values. 

In Fig.1, we show a contour plot of $^6$Li abundance as a function of
 the lifetime and the number density of CHAMPs. Since the CBBN reaction rate
 $\text{D}((C,^4\text{He}),C)^6\text{Li}$ rapidly decreases due to the 
Coulomb suppression, most of $^6$Li production occurs from $T=$ 10keV to
 a few keV. For a CHAMP lifetime $\tau_C>10^5$ sec, the $^6$Li abundance is 
insensitive to $\tau_C$. Also we found that the CBBN does not change D, $^4$He 
abundances much through $\text{D}((C,^4\text{He}),C)^6\text{Li}$. 
\begin{figure}[tbp]
\postscript{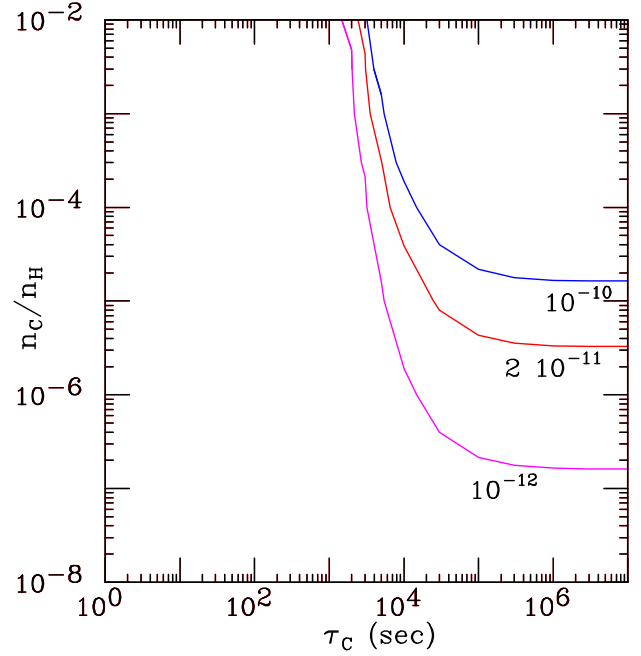}{0.95}
\caption{Contour plot for $^6$Li/H=$10^{-10}$, $2\times 10^{-11}$, $10^{-12}$ 
as a function of the lifetime and relic number density of CHAMPs 
\label{fig:photonsummary} }
\end{figure}

In the last part of this section, we will make a few remarks. 
First, since CHAMPs would be metastable, high energy injections at late
 time due to the decays are expected and they could change the value 
of several light element abundances. But notice that the amount of the energy 
injection is in princeple independent from the number density which 
can be constrained by our discussion. Of course, in specific models, 
the effects may not be negligible. On the other hand, since the current good 
agreement with observations has disfavored large changes from SBBN 
predictions for D, $^4$He, if we impose constraint on high energy injections 
from light elements other than $^6$Li, our evaluation of CBBN production of 
$^6$Li here still may provide the correct prediction. 
Of course, the late time energy injection itself also can change $^6$Li 
without any big change of the other light elements if the energy release is large, 
which typically increases $^6$Li.

The most stringent constraint on hadronic energy injections from 
several light elements is 
$\xi=Br_h\epsilon(n_C/n_{\gamma})<2\times 10^{-13}\text{GeV}$ 
at $t=10^5$sec~\cite{hadronicBBN:2004qu}, 
where $\epsilon$ is the injected energy due to 
the decay and $Br_h$ is the hadronic branching ratio of the decay. 
Roughly the above constraint implies that 
$n_C/n_H< 10^{-2}(100\text{GeV}/\epsilon)(Br_h/10^{-3})^{-1}$. 
Similarly, for $\tau_C>10^7$sec, electromagnetic(EM) energy injections obtain 
$\xi=Br_{\text{EM}}\epsilon(n_C/n_{\gamma})< 6\times 10^{-13}$GeV from 
$^6$Li/H~\cite{hadronicBBN:2004qu}, where $Br_{\text{EM}}$ is the EM branching 
ratio of the decay, that is, $n_C/n_H< 3\times 10^{-5}(100\text{GeV}/\epsilon)
(Br_{\text{EM}}/1.0)^{-1}$. The EM constraints 
significantly weaken for $\tau_C<10^7$sec. 
We do not expect a significant change for $^6$Li abundance from our 
estimate in the region of Fig.1 if $\epsilon$ is not large.  
\footnote{The direct destruction of bound $^4$He by CHAMP decays
~\cite{Kaplinghat:2006qr} may provide additonal contributions to 
$^6$Li from $Q^2<(1\text{GeV})^2$ in $\tau_C\sim10^5$sec. But the discussion 
 in $Q^2<(1\text{GeV})^2$ is still uncertain. 
Also if the mass between CHAMPs and the decay product 
highly degenerate $\sim O(m_{\pi})$ and the main decay mode happens through 
a coupling with a W-boson, new decay modes of bound states may have to be 
considered~\cite{Jittoh:2007fr}. The destruction of $^6$Li through this new 
decay mode can be important only if the capture of $^6$Li process is 
efficient at $T< 10$ keV. In this paper, these contributions
 have not been included. } 

Next, if there is no significant charge asymmetry in the CHAMP sector, 
 the positively charged and negatively charged CHAMPs can annihilate into
 two photons, W/Z or other SM particles. The energy release due to
 annihilation at $T\sim$10keV was estimated as follow~\cite{Jedamzik:2004er,
Bird:2007ge}:
\begin{eqnarray}
\xi\sim 5\times 10^{-13}\text{GeV}
(\frac{100\text{GeV}}{m_C})^{1/2}(\frac{\frac{n_C}{n_H}}{0.01})^2
\end{eqnarray}  
Again comparing with $\xi < 2\times 10^{-13}\text{GeV}(Br_h/1.0)^{-1}$, 
this effect would be irrelevant in most of the region of Fig.1.

\section{Connecting $^6$Li with the Reheating Temperature}
In our scenario, the primordial $^6$Li abundance depends 
sensitively on the reheating temperature. 
In Fig.2, assuming $\tau_C>10^4$sec, we show the contour plot for $\Omega_Ch^2=0.1$ 
which corresponds to the present dark matter relic density, 
and the contour plot for $n_C/n_H=3\times 10^{-6}$ where 
the number density of CHAMPs with $\tau_C>10^6$sec may explain the 
observed $^6$Li abundance $^6\text{Li}/\text{H}\simeq 2\times 10^{-11}$. 
It is well-known that the SBBN prediction 
of primordial $^6$Li abundance$\sim O(10^{-14})$ is much below the observed 
value. For long-lived CHAMP with $\tau_C>10^6$sec, the $^6$Li/H line 
may be a solution to explain the observed abundance of $^6$Li by 
the primordial origin. For the region above the line, 
$^6$Li may be over-produced and potentially can constrain the reheating 
temperature for $\tau_C>10^6$sec. 
On the other hand, it will require the knowledge of the late time evolution 
of $^6$Li from the primordial era to the present time.
As we can see in Fig.1, if the lifetime is shorter than $10^4$sec, 
the constraint significantly weakens. A discovery of CHAMPs with
 lifetime longer than $10^4$sec at a collider may indicate reheating beyond 
the standard scenario and put an upper bound of the reheating temperature for the models
 we are considering in this paper. 

\begin{figure}[tbp]
\postscript{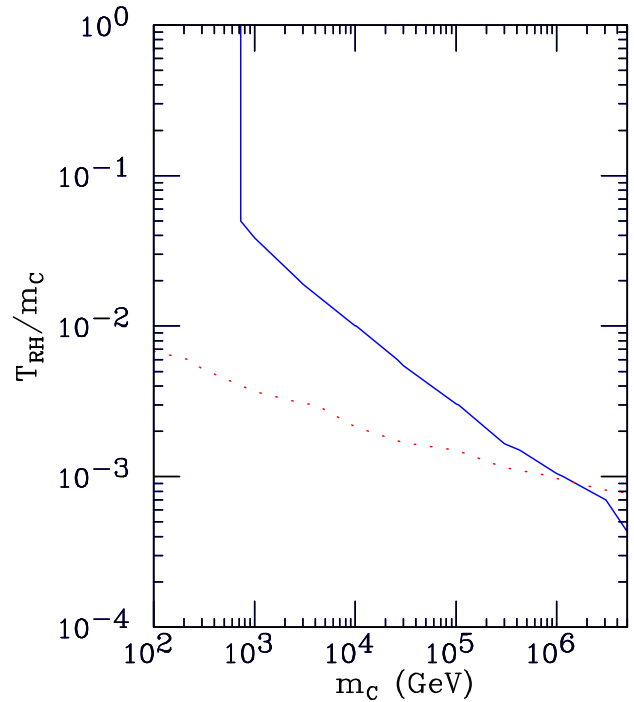}{0.95}
\caption{Contour plot for the energy density 
$\Omega_Ch^2=0.1$ (blue solid line) and the number density of CHAMPs 
$n_C/n_H=3\times 10^{-6}$ (red dotted line) as a function of mass of 
CHAMPs and reheating temperature assuming a specific reheating model of 
the Universe. $g_{\ast}(T_{\text{RH}})/g_{\ast}(T_F)^{1/2}=\sqrt{3.4}$
 were assumed.
\label{fig:photonsummary} }
\end{figure}

Several collider bounds on the mass of CHAMPs exist. 
On the other hand, the current most stringent constraint for 
CHAMP not decaying inside a detector is 98 GeV at 95$\%$ 
C.L. from CERN LEP at $\sqrt{s}=209$GeV~\cite{Yao:2006px}. 
Currently there is no model-independent constraint through Drell-Yan 
processes at the Tevatron run-II. In the LHC, the bound may be extended up to 
700GeV for 100$\text{fb}^{-1}$ for an extremely long-lived 
charged slepton in the supersymmetric standard model~\cite{LHC}. 
For the mass regions within collider reach, the $^6$Li line suggests
 lower reheating temperature $T_{\text{RH}}\sim$ a few GeV if $\tau_C>10^4$ sec. 
The realization of such a low reheating temperature may be a challenge for 
theoretical models after inflation.
\footnote{The discussions on models that such low reheating 
is generic may be interesting~\cite{future}.}  

For the mass region above the collider reach, if such heavy CHAMP decays into
 a highly degenerate neutral particle, the relic density of present dark matter 
may link to the relic density of the CHAMPs like superWIMP dark matter scenario~\cite{superWIMP}
 and still their trace may appear in astrophysical observables. 
If heavy extremely long-lived CHAMP keeps the tight coupling with
 baryon-photon fluid until it decays, the small scale power of 
primordial fluctuation of CHAMPs would be erased within the scale of
 the sound horizon at the decay time, which has an impact on the small scale
 structure formation of the Universe~\cite{CHAMPstruc}.
To explain small scale problems like missing satellite problem, which may require damping of 
the power spectrum of dark matter at small scales, the lifetime $\sim 10^7$sec is preferred, 
where the CBBN $^6$Li production may be expected.
\footnote{The rough estimation of the cut off scale is 
$\lambda\sim 0.265\text{Mpc}\sqrt{\tau_C/\text{year}}$
.\cite{Jedamzik:2005sx} Also the capture by 
$^4$He will not disturb the discussion because the bound state would 
keep similar tight coupling with baryon-photon fluid to freely 
propagating case. Also since $n_C<n_{^4\text{He}}$, the most of 
CHAMPs is captured by $^4$He, so the conclusion may not be largely different
 from the case of no-capture.}
To avoid large free streaming effects and additional cosmological 
constraints such as the cosmic microwave background black-body spectrum, tiny mass degeneracy may be 
favored, which means $\Omega_Ch^2\sim 0.1$ is the interesting region. 
We can find that such an interesting spot appears for $m_C\sim 10^6$GeV and 
$T_{\text{RH}}/m_C\sim 10^{-3}$. 
The realization of the thermal leptogenesis scenario~\cite{Fukugita:1986hr} may be 
possible if the maximum temperature is high enough to go through 
 an electro-weak phase transition, which depends on initial energy density of 
the $\phi$ field.

\section{Conclusion}
We considered extremely long-lived CHAMPs as a probe of the early Universe. 
 Such CHAMPs could leave their trace in astrophysical observables. 
The primordial $^6$Li abundance in CBBN could be sensitive 
to the number density of extremely long-lived CHAMPs in the cosmic time $t\sim 10^{3-4}$ sec, and
 if the reheating temperature of the Universe is below the mass of the CHAMPs, in a model we considered
 in this paer, the relic abundance of the CHAMPs is also sensitive to the reheating temperature.
 That is, the observed $^6$Li may indicate some information of the nature of reheating of the Universe.
\footnote{In the similar way, we also could consider the case 
of particle gravitational production due to expansion of the Universe during 
inflation~\cite{Chung:2001cb}.}

If the CHAMP mass is within the future collider reach, the measurement of the decay properties 
and the lifetime of CHAMPs are interesting challenges in future experiments~\cite{TrapCollider}. 
Since the lifetime and several parameters associated with the relic abundance may be measured 
at collider experiments, by use of the inputs, we will be able to estimate the relic density 
of CHAMPs as a function of the reheating temperature and the effects on light element abundances. 
The further understanding of CBBN and late time evolution of $^6$Li 
from the primordial era to the present time will help to understand the reheating nature of the Universe
 which may be a hidden part in the collider experiment itself.

Even if the mass range is outside of the collider reach, still 
the existence of CHAMPs may leave a trace on the power spectrum of the 
primordial fluctuation if the CHAMP was the parent particle of the present 
superWIMP dark matter~\cite{CHAMPstruc}.

 As one of the attractive possibilities, 
the search for the superWIMP sector, e.g., in supersymmetric theory, is an 
interesting target in future collider experiments and cosmological 
observations.

\section{Acknowledgements}
F.T would like to thank Csaba Csaki, Maxim Perelstein 
for reading the draft of this paper and making useful comments. 
This work is supported in part by the NSF Grant No. PHY-0355005.


\end{document}